\documentclass[apj]{emulateapj}
\usepackage{amsmath, amsthm, amssymb}

\newcommand{\etal}{{\sl{}et~al.}}

\shortauthors{Budav\'ari et~al.}
\shorttitle{Angular Clustering with Photometric Redshifts}
\journalinfo{To Appear in The Astrophysical Journal}
\submitted{}
\begin{document}

\title{Angular Clustering with Photometric Redshifts in the Sloan Digital Sky
Survey: Bimodality in the Clustering Properties of Galaxies}

\author{Tam\'as  Budav\'ari\altaffilmark{1}, 
Andrew J. Connolly\altaffilmark{2},
Alexander S. Szalay\altaffilmark{1},
Istv\'an Szapudi\altaffilmark{3},
Istv\'an Csabai\altaffilmark{4},
Ryan Scranton\altaffilmark{2}, 
Neta A. Bahcall\altaffilmark{5},
Jon Brinkmann\altaffilmark{6},
Daniel J. Eisenstein\altaffilmark{7},
Joshua A. Frieman\altaffilmark{8,9},
Masataka Fukugita\altaffilmark{10},
James E. Gunn\altaffilmark{5},
David Johnston\altaffilmark{8,9},
Stephen Kent\altaffilmark{9},
Jon N. Loveday\altaffilmark{11},
Robert H. Lupton\altaffilmark{5},
Max Tegmark\altaffilmark{12},
Aniruddha R. Thakar\altaffilmark{1}, 
Brian Yanny\altaffilmark{9},
Donald G. York\altaffilmark{8},
Idit Zehavi\altaffilmark{8}
}

\altaffiltext{1}{Department of Physics and Astronomy, The Johns Hopkins
University, 3701 San Martin Drive, Baltimore, MD 21218, USA
\label{JHU}}

\altaffiltext{2}{Department of Physics and Astronomy, University of
Pittsburgh, Pittsburgh, PA 15260, USA
\label{Pittsburgh}}

\altaffiltext{3}{Institute for Astronomy, University of Hawaii, 
Honolulu, HI 96822, USA
\label{Hawaii}}

\altaffiltext{4}{Department of Physics, E\"{o}tv\"{o}s University,
Budapest, Pf.\ 32, Hungary, H-1518
\label{Eotvos}}

\altaffiltext{5}{Princeton University Observatory, Princeton, NJ 08544, USA
\label{Princeton}}

\altaffiltext{6}{Apache Point Observatory, P.O. Box 59, Sunspot, NM 88349,
USA 
\label{APO}}

\altaffiltext{7}{Steward Observatory, 933 N. Cherry Ave., Tucson, AZ 85721
\label{Steward}}

\altaffiltext{8}{Astronomy and Astrophysics Department, University of
Chicago, Chicago, IL 60637, USA
\label{UChicago}}

\altaffiltext{9}{Fermi National Accelerator Labroratory, P.O. Box 500,
Batavia, IL 60510, USA
\label{Fermilab}}

\altaffiltext{10}{Institute for Cosmic Ray Research, University of Tokyo,
5-1-5 Kashiwa, Kashiwa City, Chiba 277-8582, Japan
\label{UTokyo}}

\altaffiltext{11}{Astronomy Centre, University of Sussex, Falmer, Brighton,
  BN1 9QJ, UK
\label{USussex}}

\altaffiltext{12}{Department of Physics, University of Pennsylvania,
Philadelphia, PA 19101, USA
\label{UPenn}}

\begin{abstract} 

Understanding the clustering of galaxies has long been a goal of modern
observational cosmology.
Redshift surveys have been used to measure the correlation length 
as a function of luminosity and color. However,
when subdividing the catalogs into multiple subsets, the errors increase
rapidly.
Angular clustering in magnitude-limited photometric surveys has the advantage
of much larger catalogs, but suffers from a dilution of the clustering signal
due to the broad radial distribution of the sample.  Also, up to now it has
not been possible to select uniform subsamples based on physical parameters,
like luminosity and rest-frame color.
Utilizing our photometric redshift technique a volume limited sample
($0.1\!\!<\!\!z\!\!<\!\!0.3$) containing more than 2 million galaxies is
constructed from the SDSS galaxy catalog.  In the largest such analysis to
date, we study the angular clustering as a function of luminosity and
spectral type.  Using Limber's equation we calculate the clustering length
for the full data set as $r_0=5.77\pm{}0.10\,h^{-1}\text{Mpc}$. We find that
$r_0$ increases with luminosity by a factor of 1.6 over the sampled
luminosity range, in agreement with previous redshift surveys. We also find
that both the clustering length and the slope of the correlation function
depend on the galaxy type.  In particular, by splitting the galaxies in four
groups by their rest-frame type we find a bimodal behavior in their
clustering properties.
Galaxies with spectral types similar to elliptical galaxies have a
correlation length of $6.59\pm{}0.17\,h^{-1}\text{Mpc}$ and a slope of the
angular correlation function of $0.96\pm{}0.05$ while blue galaxies have a
clustering length of $4.51\pm{}0.19\,h^{-1}\text{Mpc}$ and a slope of
$0.68\pm{}0.09$. 
The two intermediate color groups behave like their more extreme `siblings',
rather than showing a gradual transition in slope.  We discuss these
correlations in the context of current cosmological models for structure
formation.

\end{abstract}


\section{Introduction} \label{sec:intro}

One of the primary tools for studying the evolution and formation of
structure within the universe has been the angular correlation
function \citep{totsujikihara}. Possibly the simplest of these
point process statistics is the angular 2-point function which
measures the excess number of pairs of galaxies, as a function of
separation, when compared to a random distribution. If the universe
can be described by a Gaussian random process then the 2-point
function will fully describe the clustering of galaxies. While this is
clearly not the case and higher order correlation functions play a
significant role in understanding the clustering of structure and its
evolution, the 2-point function remains an important statistical
tool. In this paper we will utilize the angular 2-point function to
determine the type and luminosity dependence of the clustering of
galaxies within the Sloan Digital Sky Survey \citep{york}.

Studying the angular correlation function has a natural advantage over
the spatial or redshift correlation function. By only requiring
positional information we can derive the clustering signal from
photometric surveys alone (i.e.\ without requiring spectroscopic
followup observations). Given the relative efficiencies of photometric
surveys over their spectroscopic counterparts, this enables the
correlation function to be estimated for wide-angle surveys covering
statistically representative volumes of the universe and without being
limited by discreteness error. The disadvantage of the angular correlation
function has been that it is the projection of the spatial correlation
function over the redshift distribution of the galaxy sample. For
bright magnitude limits the redshift distribution is well known (and
relatively narrow) and therefore deprojecting the angular clustering
to estimate the clustering length is relatively straightforward. At
fainter magnitudes the redshift distribution becomes broader and the
details of the clustering signal can be washed out.

We can overcome many of the disadvantages of the angular clustering if
we utilize photometric redshifts. Photometric redshifts provide a
statistical estimate of the redshift, luminosity and type of a galaxy
based on its broadband colors. As we can control the redshift interval
from which we select the galaxies (and the distribution of galaxy
types and luminosities) we can determine how the clustering signal
evolves with redshift and invert it accurately to estimate the real
space clustering length, $r_0$, for galaxies. The ability to utilize
large, multicolor photometric surveys as opposed to the smaller
spectroscopic samples means that we can subdivide the galaxy
distributions by luminosity and type without being limited by the size
of the resulting subsample (i.e.\ most of our analyzes will not be
limited by shot noise). As we expect the dependence of the clustering
signal to vary smoothly with luminosity, type and redshift, it is not
expected that the statistical uncertainties in the redshift estimates
will significantly bias our resulting measures.  

The utility of photometric redshifts for measuring the clustering of galaxies
as a function of redshift and type has been recognized when studying high
redshift galaxies \citep{connolly98, maglio99, 1999MNRAS.310..540A,
1999MNRAS.305..151R, brunner00, 2001ApJ...548..127T, 2002MNRAS.332..617F}.
In this paper we focus on studying the angular clustering of intermediate
redshift galaxies $z<0.3$ and the dependence of the clustering length on
luminosity and galaxy type. This represents one of the first applications of
the photometric redshifts to the clustering of intermediate redshift galaxies
for which we have a large, homogeneously and statistically significant sample
of galaxies. This paper is divided into five sections.
In Section 2 we describe the data set used in this analysis and the selection
of a volume limited sample of galaxies. In Section 3 we apply a novel
approach for estimating the 2-point angular correlation function using Fast
Fourier Transforms and we show the dependence of the slope and amplitude of
the correlation function on luminosity and galaxy type. In Section 4 we
invert the projected angular correlation function and derive the correlation
lengths. In Section 5 we discuss the bimodal behavior of clustering
properties.

\section{Defining a Photometric Sample}

The SDSS is a photometric and spectroscopic survey designed to map the
distribution of stars and galaxies in the local and intermediate redshift
universe \citep{york}. On completion the SDSS will have observed the majority
of the northern sky ($\sim \pi$ steradians) and approximately 1000 square
degrees in the southern hemisphere. These observations are undertaken in a
drift-scan mode where a dedicated 2.5m telescope scans along great circles,
imaging 2.5 degree wide stripes of the sky. The imaging data is taken using a
mosaic camera \citep{gunn} through the five photometric passbands $u'$, $g'$,
$r'$, $i'$ and $z'$ (covering the ultraviolet through to the near infrared)
as defined in \citet{fukugita}. The photometric system is described in detail
by \citet{smith02} and the photometric monitor by \citet{hogg01}.
All data are reduced by an automated software pipeline \citep{lupton,pier02}
and the outputs loaded in to a commercial SQL database. In our analysis, we
will include data from runs with the longest contiguous scans (typically in
excess of 50 degrees). These data comprise a subset of the data that will be
released to the public as part of data release \citep{dr1}. In comparison to
the Early Data Release \citep{stoughton} the area analyzed in this paper is
approximately $\sim\!\!10$ times larger, or approximately 20\% of the entire
survey area.

Given the five band photometry from the SDSS imaging data we estimate the
photometric redshifts of the galaxies using the techniques outlined in
\citet{budavari00, budavari99, csabai00, connolly99}. The details of the
estimation techniques employed together with the expected uncertainties
within the redshift estimators are given in \citet{csabai02}. In this paper
we will just note the effective rms error of the photometric redshifts
(typically $\Delta{}z_{\rm{}rms} = 0.04$ at $r^*<18$). For all sources within
a sample the redshift and its uncertainty are calculated together with a
measure of the spectral type of the galaxy and its variance and covariance
(with redshift). From these measures we estimate the luminosity distance to
each galaxy and calculate its r-band absolute magnitude.

\begin{figure*}
\plotone{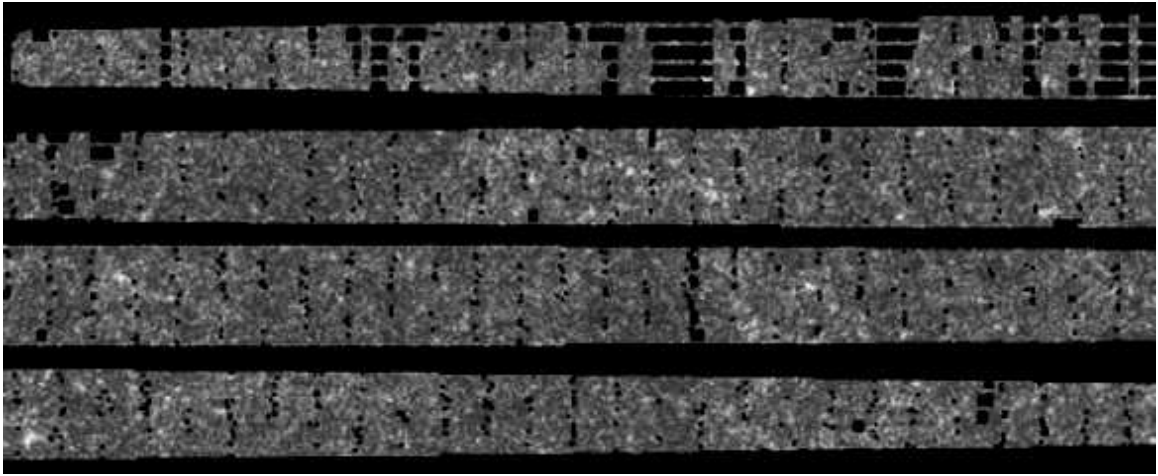}
\caption{Density of galaxies in stripe 11. The image of the stripe is split
in 4 to preserve the aspect ratio. The width of the stripe is 2.5 degrees.
The black pixels represent regions containing no objects. They are masked out
due to bad field quality, seeing or bright stars. The narrowing of the stripe
towards the left and right edge is due to the SDSS stripe geometry which
ensures that objects are not counted twice. The pieces of the rectangle
masked out towards the ends of the stripes are covered by adjacent stripes.}
\label{fig:binned}
\end{figure*}

\subsection{Building the Sample Database}

From the current photometric data in the SDSS archives we extract eight
stripes for our analysis. These stripes combine to form approximately
five coherent regions on the sky which range from approximately 90
degrees to about 120 degrees in length. In the nomenclature of the
SDSS these stripes are designated the numbers 10--12, 35--37 and 76 and
86. The last two stripes come from the southern component of the
survey. All data from these stripes have been designated as having
survey quality photometric observations and astrometry. In total these
stripes add up close to 20 million galaxies and are accessible through
the SDSS Science Archive \citep{sx}.

Currently, there are two versions of the SDSS science archive running
at Fermilab and remotely accessible to the collaboration. The
``chunk'' database contains stripes that have passed through the
target selection process for identifying candidates for spectroscopic
followup but with only photometry that was available when the
spectroscopic target selection was run on the region (i.e.\ photometry
for which the calibrations were not necessarily optimal). The
``staging'' database has the latest photometric data (for this paper
we use the {\em{}ver.~5.2.8} of the photometric pipeline) but without
the full target selection information. For our purposes the quality of
the photometric measurements is important but the target
selection is completely irrelevant and thus we take our sample from the
staging database.

In order to be able to efficiently store, search and select galaxies
from the catalog, we create a local database using Microsoft's SQL
Server. The relevant properties (position, redshift, galaxy type,
absolute and apparent magnitudes) of all galaxies in the staging
database were stored in this database server. The regions of the sky
surveyed by these data, the seeing of the observations as a function
of position on the sky and the position of bright stars that must be
masked out when defining the survey geometry were all calculated
internally from this data set.

From these data we restrict our sample to galaxies brighter than
$r^*=21$. At this magnitude limit the star-galaxy separation is
sufficiently accurate that it will not affect the angular clustering
measures \citep{scran03} and the photometric redshift errors are
typically less than $\sigma=0.06$ \citep{csabai02}. Applying this
magnitude limit yields approximately 13 million galaxies from which to
estimate the clustering signal. The sample was further restricted by
excluding those regions of the stripes affected by the wings of bright
stars or that were observed with poor seeing. Figure~\ref{fig:binned}
shows the density of galaxies in stripe 11 with the masks
over-plotted. Fields with seeing worse than $1.7''$ and a
$3'\!\!\times{}\!3'$ neighborhood around all bright stars with
$r^*<14$ were discarded.

We note that these selections were all accomplished by applying a
series of SQL queries to the database rather than progressively
pruning a catalog of galaxies. The boundaries of the stripes, the
seeing on a field-by-field basis and all bright stars were stored in
the local database, so that masks could be generated on the fly over
the area to be analyzed. As such the selection criteria that were
applied to the database could be optimized in a relatively short
period of time.

\subsection{Clustering from a Volume Limited Sample}

Often the clustering evolution of galaxies, particularly that defined by
angular clustering studies, is characterized as a function of limiting
magnitude. While observationally this is simple to determine, the results of
these analyzes are often difficult to interpret because in a magnitude
limited sample the mix of the spectral types and absolute luminosities of
galaxies is redshift dependent. We are, essentially, looking at the
clustering properties of different types of galaxies as a function of
limiting magnitude. The models to account for the clustering signal must,
therefore, also be able to describe the evolution of the distribution of
galaxies. Ideally we would separate out the effects of population mixing and
study the evolution of angular clustering in terms of the intrinsic
properties of galaxies (i.e.\ rest-frame color and luminosity), along with
their distances. We can accomplish this if we use the photometric redshifts
to select a volume limited sample of galaxies (i.e.\ one with a fixed
absolute magnitude range as a function of redshift).

The SDSS Early Data Release photometric redshift catalog by
\citet{csabai02} is based on techniques \citep{budavari00, budavari99,
csabai00, connolly99} that estimate the physical parameters describing
the galaxy samples in a self-consistent way. The relationship for the
EDR data set is applied to the photometric data selected from the 
imaging stripes and the spectral types, absolute magnitudes and
k-corrections are stored within a database together with the
photometric and positional information. All derived quantities assume
an $\Lambda$ cosmology with $h=0.7$, $\Omega_{\rm{}M}=0.3$ and
$\Omega_{\Lambda}=0.7$.

\begin{figure}[b]
\plotone{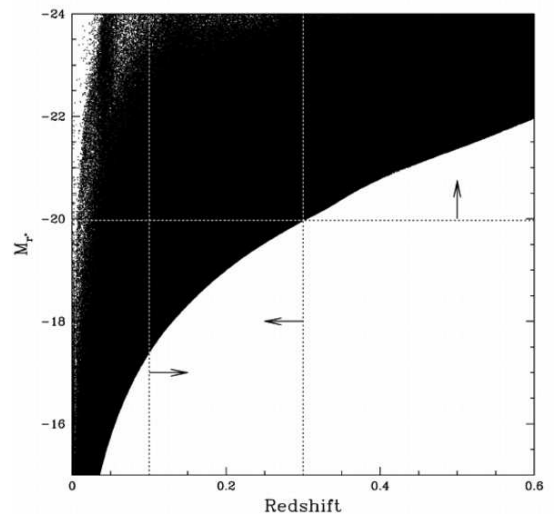}
\caption{Absolute $r^*$ magnitude vs.\ redshift relation as derived from
the photometric redshift catalog. The dotted lines show our selection for the
volume limited sample out to redshift 0.3.}
\label{fig:absmagz}
\end{figure}

Figure~\ref{fig:absmagz} illustrates how the absolute magnitude limits
in the $r^*$ band vary as a function of redshift. These absolute
magnitude boundaries are well defined, as a function of redshift, by
the galaxies within our sample. In this paper we study a
volume-limited sample that extends out to a redshift $z=0.3$ with
limiting absolute magnitude $M_{r^*}=-19.97$. We further restrict the
data set to those galaxies more distant than redshift $z=0.1$.  The
reason for this lower redshift limit is that the main spectroscopic
galaxy sample will contain many of the low redshift galaxies, so using
photometric redshift estimates is not necessary. 
Also the fractional error in $z$ becomes large at lower redshifts due to
uncertainty in the phototometric redshifts.
The final catalog size is over 2 million galaxies.

\subsection{Rest-frame Selected Subsamples}

To analyze how angular clustering changes with luminosity
in Section~3, the volume limited sample is divided into 3
absolute magnitude bins. These subsamples represented by $M_1$, $M_2$
and $M_3$ have limiting absolute magnitudes $M_{r^*}\!>\!-21$,
$-21\!>\!M_{r^*}\!>\!-22$ and $-22\!>\!M_{r^*}\!>\!-23$
respectively. The size of these subsets decrease by approximately a
factor of two as a function of increasing luminosity.

\begin{figure}
\plotone{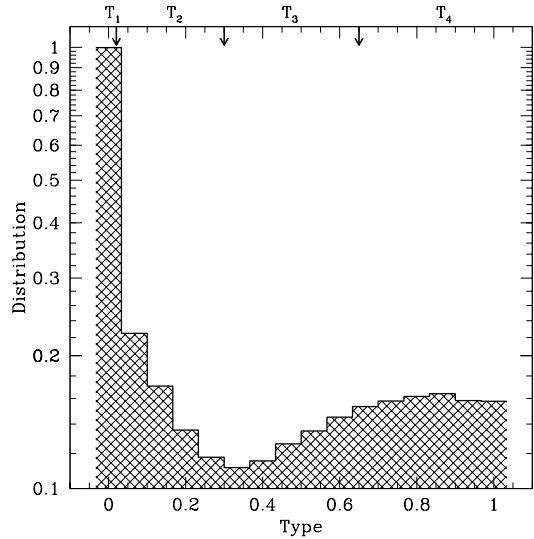}
\caption{The distribution of spectral type is bimodal. The small arrows
hanging from the top axis illustrate our subsamples selected by spectral
type. The cut between the 2 reddest and bluest classes at type $t=0.3$
(equivalent to rest-frame $u^*-r^*=2$) splits the distribution into 2 halves
of similar sizes.}
\label{fig:typehist}
\end{figure}

\begin{figure}[b]
\plotone{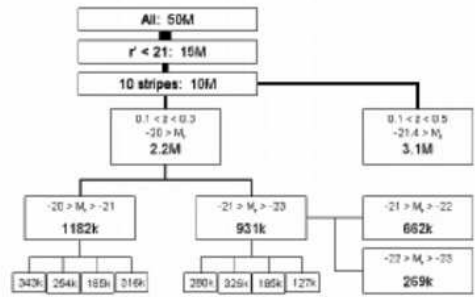}
\caption{Our classification and selection criteria are shown in this figure
along with the number of galaxies in the subsamples.}
\label{fig:select}
\end{figure}

For the type dependent selection we utilize the continuous spectral
type parameter $t$ from the photometric redshift estimation. This
essentially encodes the rest-frame colors of the galaxies as there is
a direct one-to-one mapping between the type $t$ and the spectral
energy distribution (SED) of the galaxy. The value $t=0$ represents a
template galaxy that is as red as the elliptical spectrum of
\citet{cww}; as $t$ increases the galaxy type becomes progressively later.
In
Section~3, we subdivide by spectral class breaking the
luminosity classes into four subgroups (each with comparable numbers
of galaxies). The cuts in the spectral type parameter $t$ from red to
blue are $t\!<\!0.02$, $0.02\!<\!t\!<\!0.3$, $0.3\!<\!t\!<\!0.65$ and
$t\!>\!0.65$. The cuts are defined as the $T_1$, $T_2$, $T_3$ and
$T_4$ subsamples respectively. Our selection is motivated by the
spectral energy distributions of \citet{cww}. The first class consists
of galaxies with SEDs similar to the CWW elliptical template (Ell),
the second, third and fourth classes contain a broader distribution of
galaxy types approximately corresponding to Sbc, Scd and Irr types,
respectively. The distribution of types and our classification are
shown in Figure~\ref{fig:typehist} and \ref{fig:select}.

\section{The Angular Correlation Function}

The properties of the angular correlation function and the estimators
used to measure it from photometric catalogs have been extensively
discussed in the astronomical literature \citep{kerscher00}. The probability of
finding galaxy within a solid angle $\delta\Omega$ on the celestial
plane of the sky at distance $\theta$ from a randomly chosen object is
given by \citep{peebles}
\begin{equation}
\delta{}P = n\ [1+w(\theta)]\ \delta\Omega ,
\end{equation}
where $n$ is the mean number of objects per unit solid angle. The angular
two-point correlation function $w(\theta)$ basically gives the excess
probability of finding an object compared to a uniform Poisson random point
process.

Traditionally the observations are compared to random catalogs that
match the geometry of the survey. The computation usually consists of
counting pairs of objects drawn from the actual and random catalogs
and applying a minimum variance estimator such as that defined by
\citet{landyszalay} or \citet{hamilton}. In this study, we use the
Landy-Szalay estimator \citep{landyszalay} as
\begin{equation}
w_{ LS} = \frac{\rm{}DD-2DR+RR}{\rm{}RR} ,
\label{eq:estimator}
\end{equation}
where DD, RR and DR represent a count of the data-data, random-random
and data-random pairs with $\theta$ angular separation summed over the
entire survey area.

\begin{figure}
\plotone{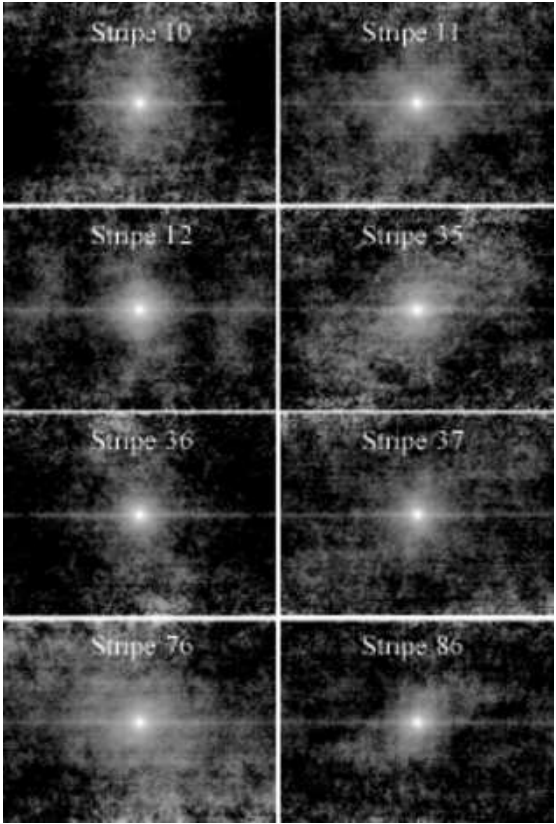}
\caption{The 2-dimensional correlation functions are shown for all 8 stripes
on logarithmic scale. The correlation function is expected to be
isotropic. Its sensitivity to artifacts in the data makes it an excellent
diagnostic tool. The horizontal direction is the scan direction, along the
stripe. The elongated streak at zero lag is related to the flat-field vector
(see text). Note that the plots have parity symmetry through the origin, thus
any half of the 2D space contains all the information for computing
$w(\theta)$.}
\label{fig:twodim}
\end{figure}

\subsection{Estimating $w(\theta)$ with a Fast Fourier Transform}

Even though distances on the sky are easy to compute mathematically,
measuring the correlation function is not a trivial task, especially
when it comes to large surveys. It's easy to see that any naive
algorithm implementing the estimator in Eq.~(\ref{eq:estimator})
scales with the square of the number of objects $N$ in the survey,
${\cal{}O}(N^2)$. It, therefore, becomes progressively more expensive
in computational power to apply these techniques to data the size of
the SDSS samples.

We propose a novel method to estimate $w(\theta)$ using the Fast Fourier
Transform \citep[e.g.~in][hereafter FFT]{nr}, which scales as
${\cal{}O}(N\log{}N)$, a significant improvement over the naive approach. The
principle behind this is to group all galaxies into small cells within a grid
and analyze this matrix instead of the point catalog. The implementation of
this FFT approach called eSpICE is a Euclidean version of SpICE by
\citet{spice}. A discussion on its properties and scalings is given elsewhere
\citep{espice}. Here we present an application of eSpICE to the SDSS
galaxy angular clustering and limit our discussion of the algorithm to only
an outline of how the method works.

To apply a standard FFT analysis to the problem we operate in Euclidean
space. In other words, Euclidean distances are computed instead of the correct
angular separation. 
Given that we only examine separations of less than 2 degrees, the maximum
relative distance error introduced by the use of the Euclidean approximation
is only 0.00005. This has a negligible effect on the accuracy of our
analysis, and the use of Euclidean distances increases the speed of the
algorithms significantly.
We use SDSS survey coordinates ($\mu,\nu$) to define the position of an
object within a stripe. These coordinates are defined locally for each
stripe. As the individual stripes comprise great circles on a sphere the
survey coordinates along these great circles are close to Euclidean (i.e.\ in
this coordinate system every stripe looks as if it were equatorial). In fact
for equatorial stripes 10 and 82, the ($\mu,\nu$) coordinates are the same as
(RA, Dec). Beyond the above {\em data} grid of galaxies, we need to describe
the geometry of the survey area. This is done by a second grid, with the same
dimensions that describes the survey boundaries. We call this the {\em
window} grid because it is 1 if the pixel is entirely inside the boundaries
and 0 otherwise. In this way we can also incorporate arbitrarily complex
masks (e.g.\ for excluding regions around bright stars) as they are simply
applied to the window grid. The data and window grids are then padded with
zeros up to the maximum angular scale to avoid aliasing in the Fast Fourier
Transform. Finally, eSpICE is used to calculate the two-point correlation
function directly.

Our approach is to apply this algorithm to one stripe at a time. The
data and window matrices are constructed by querying the SQL database
to select the appropriate galaxies and masks. From these data we
calculate the two-dimensional angular correlation function for all
eight stripes. Figure~\ref{fig:twodim} shows the 2D correlation
function for all stripes used in this analysis.  We find that the
2D correlation function is an extremely sensitive diagnostic tool for
identifying systematics within the photometric data. The correlation
function is expected to be isotropic. Artifacts within the data or
survey geometry distort this symmetry. This is seen, to varying
degrees, in each of the eight 2D correlations functions.  We find that
within the 2D correlation function there is an elongated streak, at
zero lag, along the scan direction which has structure on scales in
excess of a few degrees. This arises due to errors in the flat-field
vector. 

In a drift-scan survey the flat field is a one-dimensional vector
(orthogonal to the direction of the scan). Errors within the flat
field tend, therefore, to be correlated along the scan direction
(i.e.\ along the columns of a stripe). This effect is seen within all
of the individual stripe correlation functions. As it is, by
definition, a zero lag effect we can exclude it from the analysis by
censoring this region of the 2D correlation function before
azimuthally averaging the signal to get a 1D correlation function. We
note, however, that even if we do not censor the data to remove this
effect the result of averaging the correlation function azimuthally
(and the fact that this elongated streak only affects a small fraction
of the 2D correlation function) the results discussed in the following
sections are not affected.

\begin{figure}
\plotone{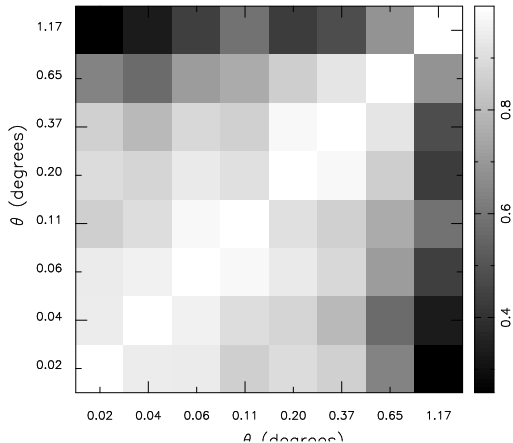}
\caption{The covariance matrix of the entire volume limited sample is
normalized so that the diagonal elements are white and the grayscale values
represent the correlations.
\label{fig:covar}}
\end{figure}

In total, the computation of the one-dimensional angular correlation function
$w(\theta)$ takes less than 3 minutes for a stripe, which is several orders
of magnitude faster than traditional two-point estimators. Having computed
the correlation function for all stripes, we co-add the results by properly
weighting with the number of galaxy pairs.  At the same time the covariance
of the signal is also estimated from the scatter between the different
stripes Figure~\ref{fig:covar} shows the covariance matrix for the volume
limited sample ($z<0.3$). The image of the matrix is normalized so that the
diagonal elements are always white and the (off-diagonal) gray scale values
represent the correlation between the bins. Because of the logarithmic
sampling of $w(\theta)$, the neighboring bins are farther apart and hence
correlate progressively less as we go to larger scales and thus to larger
bins.

\subsection{The Clustering Scale Length}

With the measured correlation amplitudes as function of angular
separation in hand, we can obtain a parametric form of the
scaling. The amplitude and power of the correlation are calculated by
fitting the usual formula,
\begin{equation}
w(\theta) = A_{w} \left(\frac{\theta}{\theta_0}\right)^{-\delta} .
\end{equation}
Normalizing $\theta$ by $\theta_0=0.1^{\circ}$ enables the amplitude $A_w$ be
directly compared with the figures showing $w(\theta)$ measurements; $A_w$ is
essentially the value of the correlation function at $\theta_0$, since $A_w
\equiv w(\theta_0)$.  The parameters $A_w$ and $\delta$ are estimated by
minimizing the cost function
\begin{equation}
\chi^2(A_{w},\delta) = \frac{1}{N\!\!-\!\!2}\sum_{i,j=1}^N 
	\Delta w_i\ C^{-1}_{ij} \Delta w_j ,
\end{equation}
where $\Delta w_i = w_i - w(\theta_i | A_w, \delta)$, $N$ is the
number of bins and $C_{ij}$ is the covariance matrix. Although,
$\chi^2(A_w,\delta)$ is not quadratic in the parameter $\delta$, it is
in $A_w$, thus the equation
\begin{equation}
\frac{\partial\chi^2}{\partial{}A_w} = 0 
\end{equation} 
can be solved analytically for $A_w$ to reduce the dimensionality of
the problem. We then use a method by \citet{brent} \citep[see][]{nr}
to search for the optimal value of $\delta$. The $\chi^2$ fit also
provides information about the covariances of the estimated parameters
that are shown as error ellipses in the next section.

\begin{figure*}
\plottwo{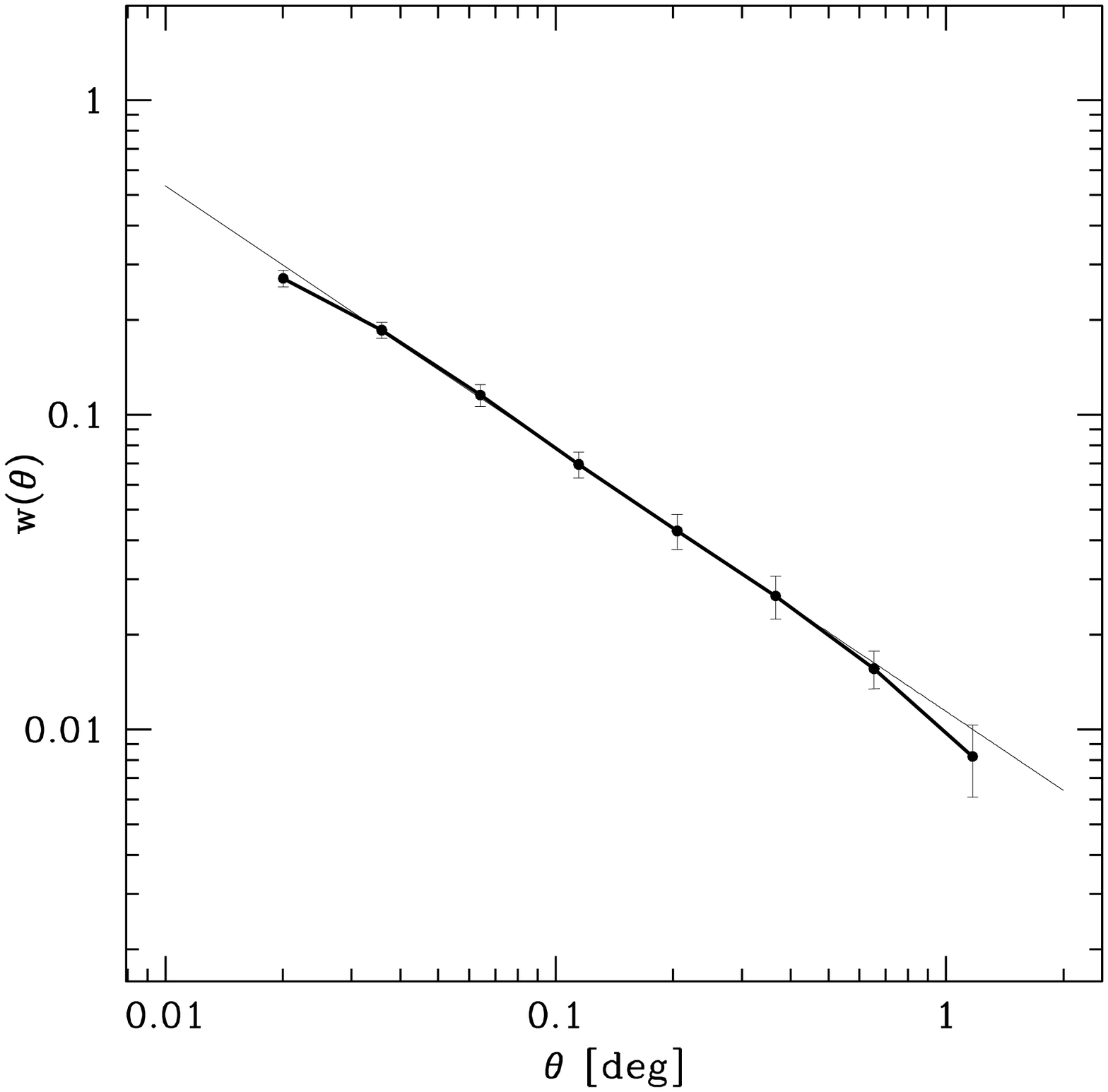}{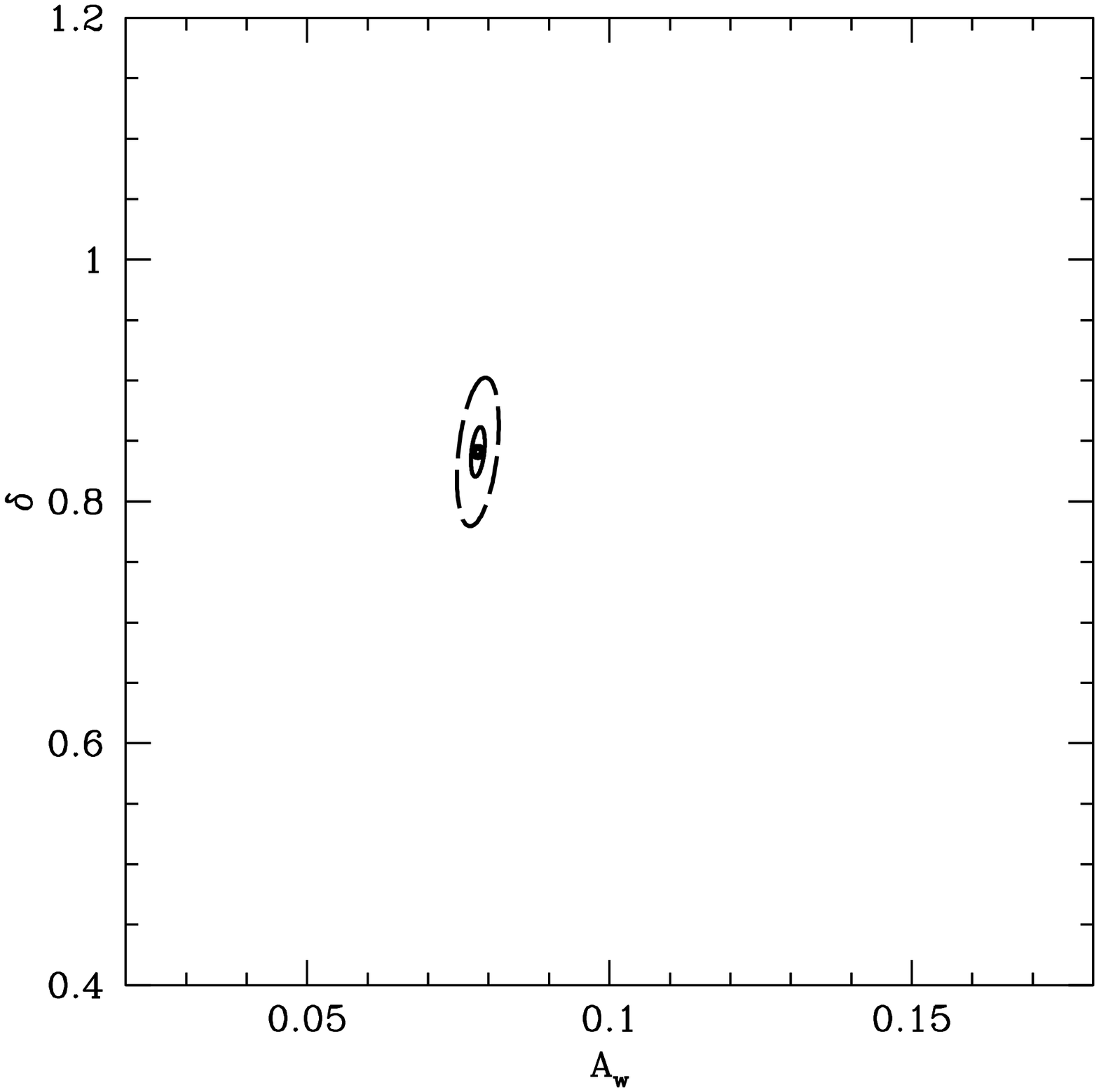}
\caption{Angular clustering in the fiducial sample of galaxies in the volume
limited sample \mbox{($z<0.3$)}. The correlation function is shown ({\it{}left
panel}) along with the best power-law fit using the formula
$w(\theta)=A_w(\theta/0.1^{\circ})^{-\delta}$ ({\it{}right panel}).}
\label{fig:fid}
\end{figure*}

Figure~\ref{fig:fid} illustrates the angular clustering in our
fiducial sample of all galaxies within the volume limited sample. In
the left panel, the correlation function $w(\theta)$ is shown along
with the best power-law fit. The range in angular separation varies
from 1 arcminute to 2 degrees. At a mean redshift for the volume-limited
sample of $z=0.2$, $2^{\circ}$ corresponds to $\sim\!17\, 
h^{-1}\rm{}Mpc$. The error bars on the measurements are computed as
one over the square root of the diagonal elements of the covariance
matrix (e.g.\ shown in Figure~\ref{fig:covar}). In the right panel of
Figure~\ref{fig:fid}, we plot the best fit parameters and their error
ellipses. For the full volume-limited sample the best fit to the data
has a slope of $\delta=0.84\pm{}0.02$ with an amplitude of
\mbox{$A_w=0.078\pm{}0.001$}.

\begin{figure*}
\plottwo{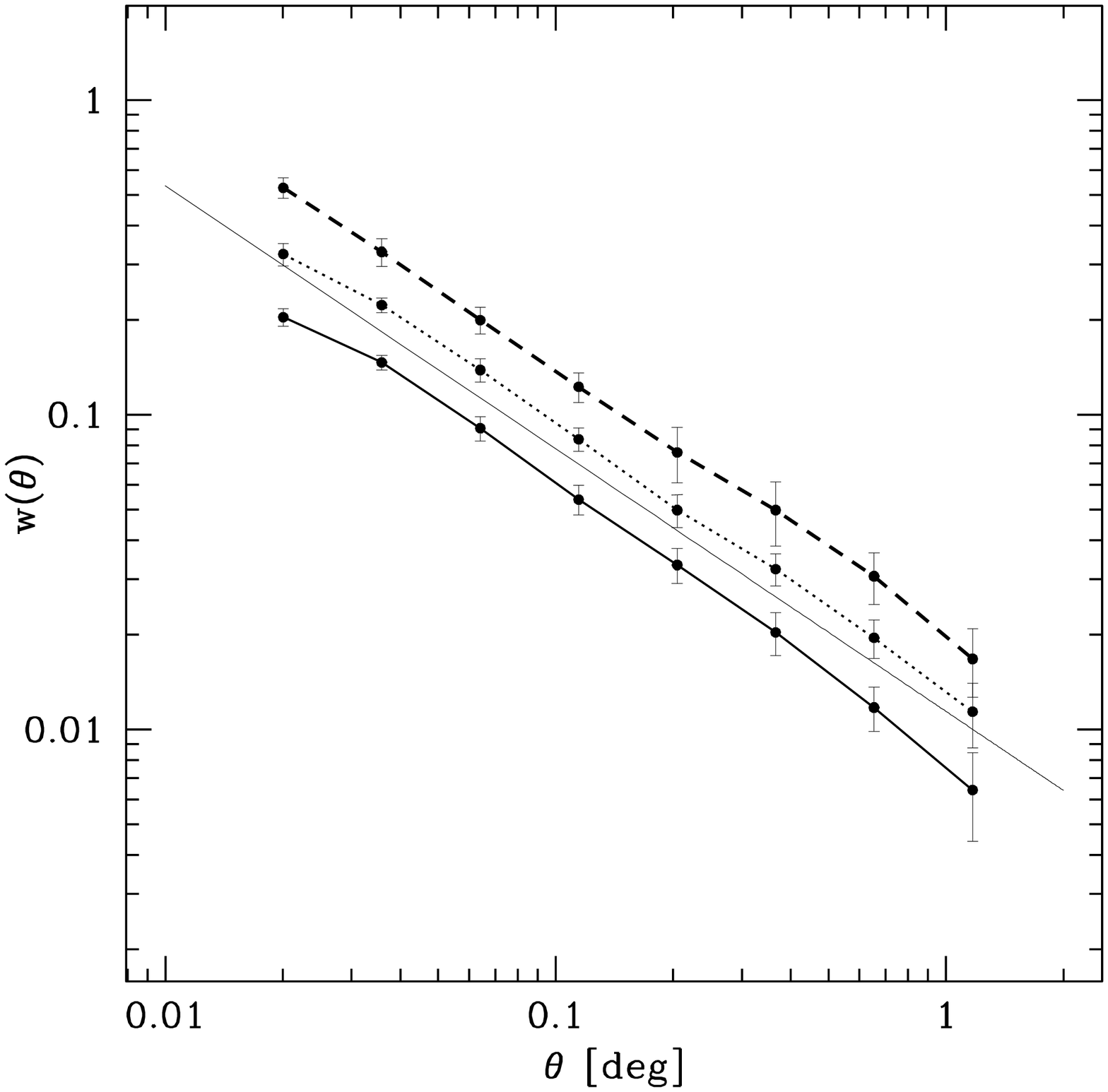}{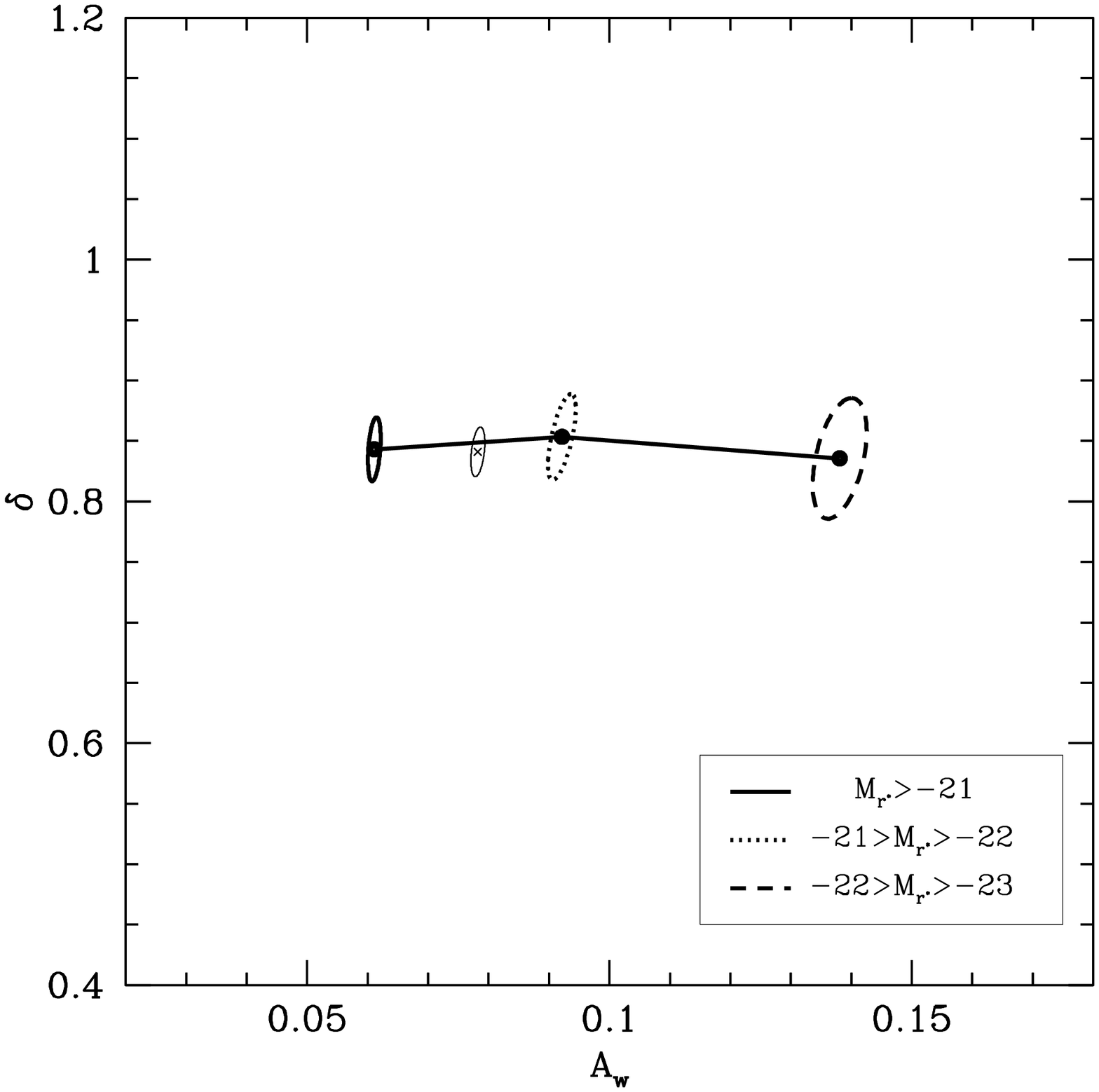}
\caption{The clustering strength changes as a function of absolute magnitude
without change in the slope. The correlation function and the parameter fits
are shown in the left and right panels, respectively. The straight line
represents the best fit to the fiducial sample in Fig.~\ref{fig:fid}.}
\label{fig:lum}
\end{figure*}

\subsection{Clustering as a Function of Luminosity}

The angular clustering of the galaxies in the three luminosity bins
described in Section 3 are compared in Figure~\ref{fig:lum}. As noted
earlier, the left panel shows the correlation function, and the right
panel gives the parameters of the power-law fits. The first section
(I) of Table~\ref{tbl:all} presents the values and errors on these
measured parameters. As expected, the more luminous galaxies are
clustered more strongly: the amplitudes are roughly larger by a factor
of 1.5 from one sample to the next and are measured to be
$0.061\pm{}0.001$, $0.092\pm{}0.002$ and $0.138\pm{}0.004$. The slope
of the correlation functions are consistent for all luminosity bins
and with the fiducial value derived for the entire volume (the
estimated slope parameters scatter around $\delta=0.84$).

\subsection{Bimodality of $w(\theta)$ as a Function of Rest-frame Color}

\begin{figure*}
\plottwo{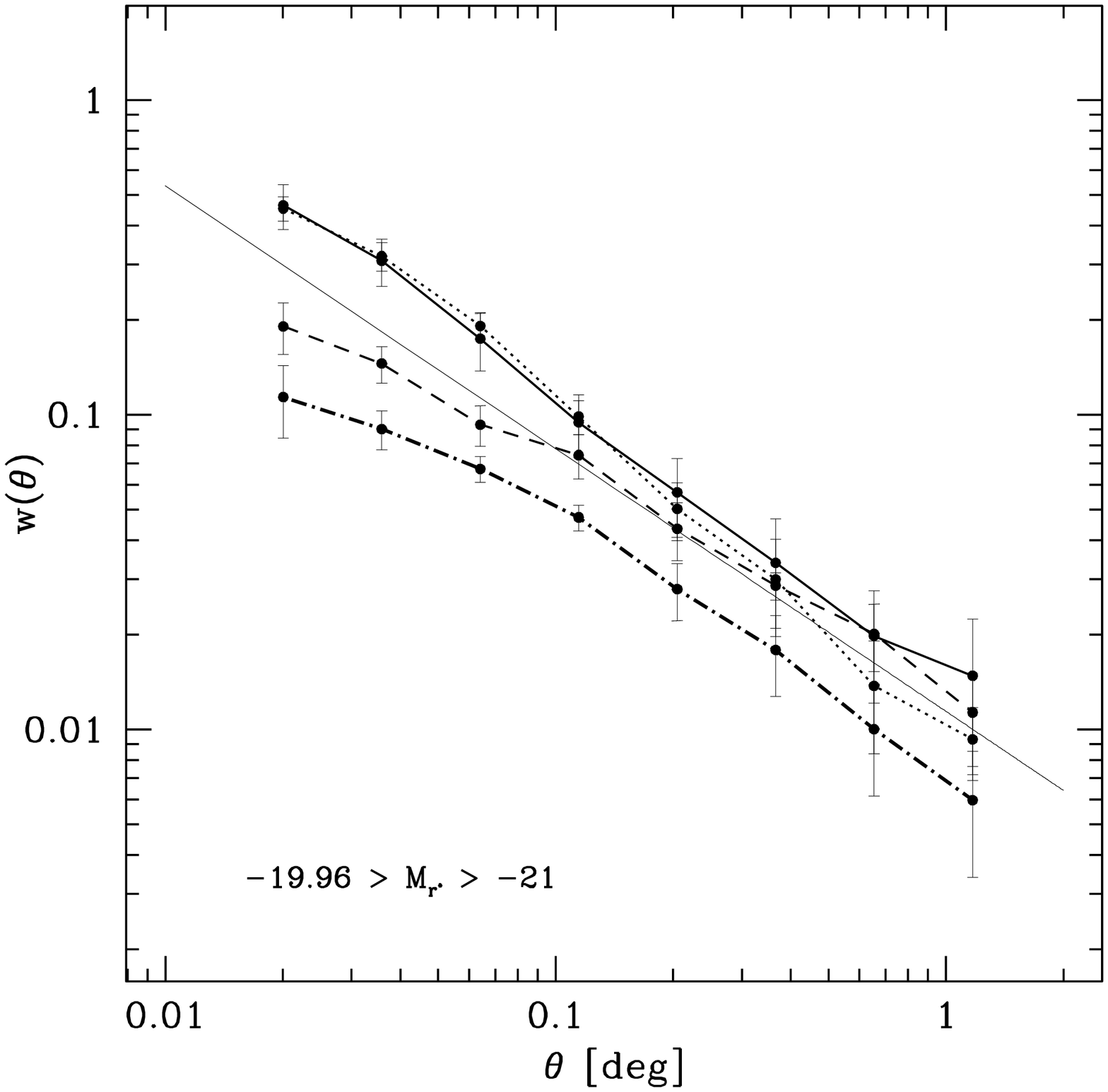}{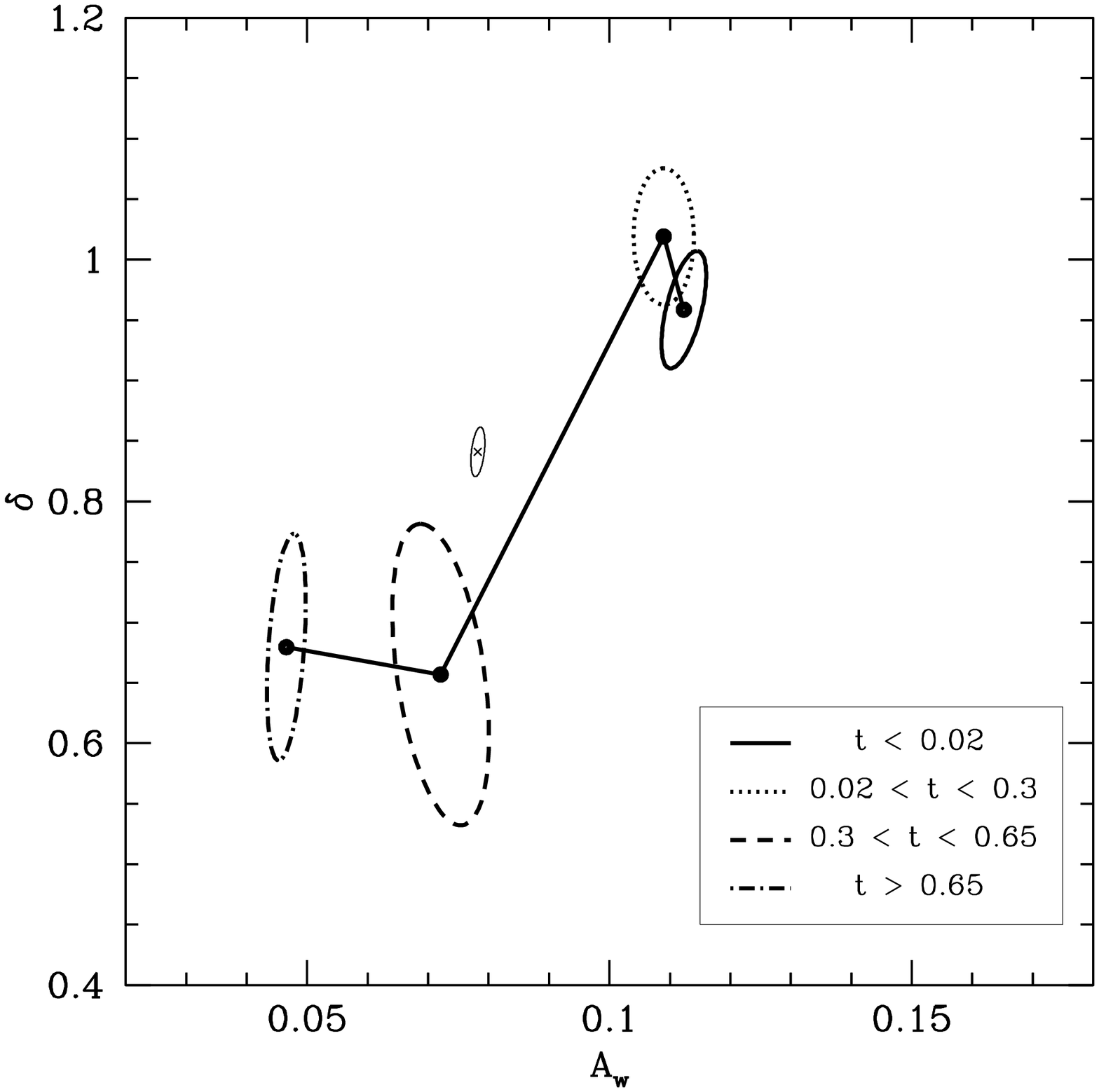}
\caption{The slope of the angular correlation function changes significantly
as a function of SED type. The figures illustrate the trend in the lower
luminosity bin with $M_{r^*}\!>\!-21$. Note the bimodality in slope: the
reddest two classes have the same slope, and the bluest two classes have the
same slope, but the red slope differs distinctly from the blue slope.}
\label{fig:lo}
\end{figure*}

\begin{figure*}
\plottwo{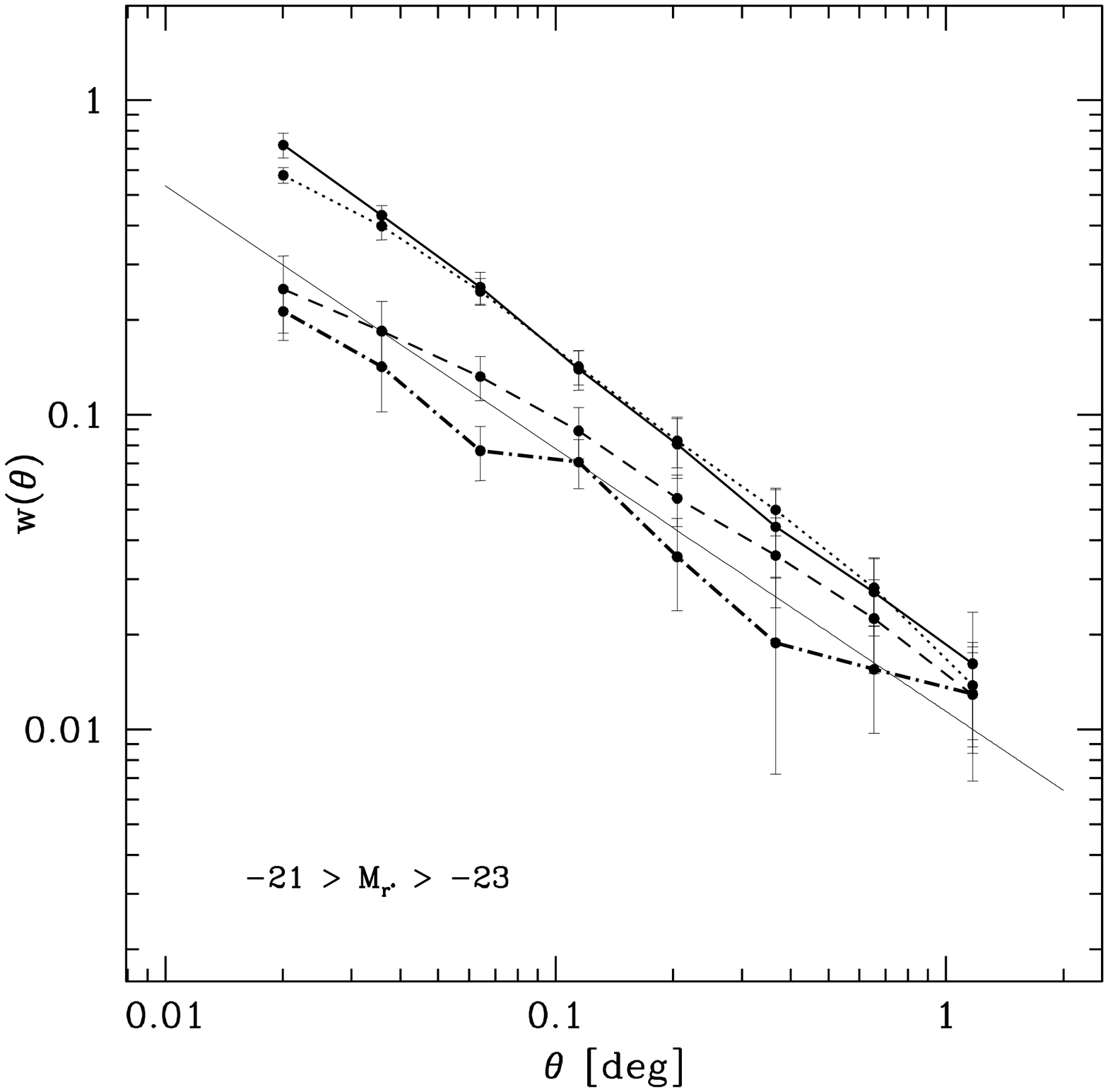}{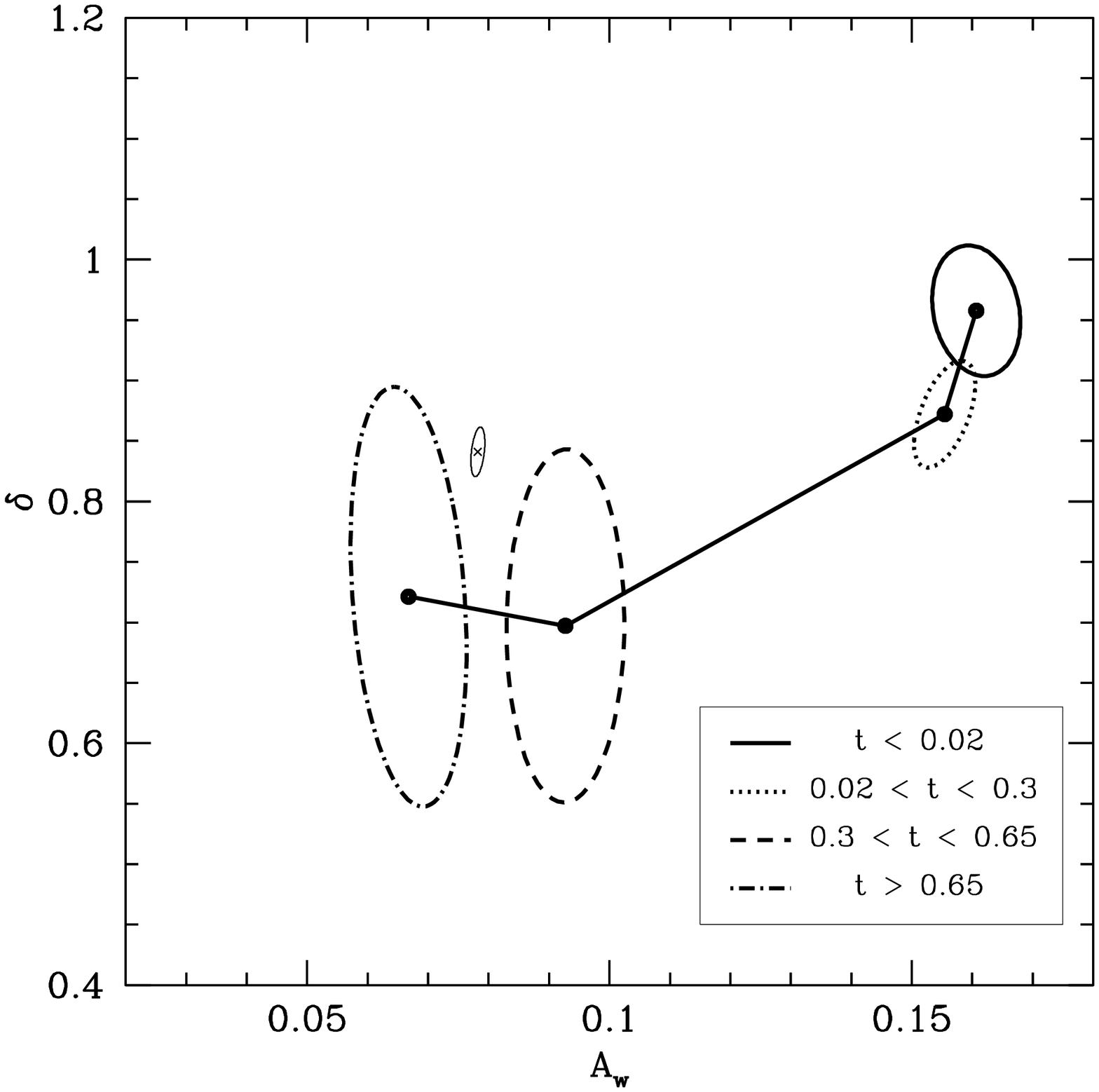}
\caption{The clustering results follow the same trend in the high luminosity
bin with $-21\!>\!M_{r^*}\!>\!-23$ as in the lower luminosity bin seen in
Figure~\ref{fig:lo} but the amplitudes are systematically larger.}
\label{fig:hi}
\end{figure*}

The type dependence of the correlation function is
not as simple as that found for the luminosity classes. It is well known that
different types of galaxies have different clustering behavior
\citep{giovanelli}. Red elliptical galaxies are more likely to be found in
higher density regions than spirals. Here, we study the evolution of the
angular correlation function with spectral type in two absolute magnitude
ranges, $M_{r^*}\!>\!-21$ and $-21\!>\!M_{r^*}\!>\!-23$. Both of these yield
approximately 1 million galaxies within our volume. Figure~\ref{fig:lo} shows
how the clustering changes as a function of spectral type in the lower
luminosity sample. The 2 reddest classes of the galaxy population ($T_1$ and
$T_2$) are essentially indistinguishable, their clustering is significantly
stronger than for the other classes or our fiducial results.  The bluest 2
classes ($T_3$ and $T_4$) have approximately the same power-law exponents but
with different amplitudes. Section II of Table~\ref{tbl:all} summarizes the
results of parameter fitting, which is also seen in the right panel of
Figure~\ref{fig:lo}. The higher luminosity classes show the same basic trends
but with a stronger correlation amplitude. The results of the type dependence
of power law fits to the high luminosity class are given in
Figure~\ref{fig:hi} and Table~\ref{tbl:all}.

\section{The Correlation Length for Galaxies}

\subsection{Limber's Equation}

From the angular clustering we can derive the spatial correlation
length, $r_0$, given the redshift distribution of the data
\citep{peebles}. This is accomplished by integrating over the comoving
coordinate along two lines of sight $r_1$ and $r_2$, separated by the
angle $\theta$, to calculate the projected angular correlations from
the real-space correlation function $\xi(r) = (r/r_0)^{-\gamma}$,
\begin{equation}
w(\theta) = \int \frac{r_1^2 \Phi(r_1) dr_1}{F(r_1)}
\int \frac{r_2^2 \Phi(r_2) dr_2}{F(r_2)}\ \xi_{12} ,
\end{equation}
where $\Phi(r)$ is the selection function and the factor $F(r)=1-kr^2$
accounts for the curvature of space.%
\footnote{In our case $F(r)$ would be constant $1$ because the parameters
  $\Omega_M=0.3$ and $\Omega_{\Lambda}=0.7$ assume a flat universe, however,
  we will do the integral in redshift (see later).
}
In the small angle approximation, the correlation function
$\xi_{12}=\xi(r_1^2+r_2^2-2r_1r_2\cos\theta)$ becomes
$\xi_{12}=\left(r/r_0\right)^{-\gamma}
\left(\theta^2+u^2/r^2\right)^{-\gamma/2}$ using variables $2r=r_1+r_2$ and
$u=r_1-r_2$. This simplifies the above integral, because one can separate out
the part that has $u$ and arrive at
\begin{equation}
w(\theta) = r_0^{\gamma} H_{\gamma} \theta^{1-\gamma}
   \int \frac{r^{5-\gamma}\Phi^2(r)}{F^2(r)} dr ,
\end{equation}
where 
\begin{equation}
H_{\gamma} = {
\frac{\Gamma(\frac{1}{2})\ \Gamma(\frac{\gamma-1}{2})}
     {\Gamma(\frac{\gamma}{2})}
} .
\end{equation}  
We can rewrite the integral using redshift $z$ and its
distribution. Substituting $r^2\frac{\Phi}{F}dr=\frac{dN}{dz}dz$ one
gets the final integral of
\begin{equation}
w(\theta) = r_0^{\gamma} H_{\gamma} \theta^{1-\gamma}
   \int_0^{\infty} r^{1-\gamma} \left(\frac{dN}{dz}\right)^2 
   \left(\frac{dr}{dz}\right)^{-1} dz
\end{equation}
that may be compared to the measured quantities directly. We have
assumed that the selection function and the redshift distribution are
normalized to $\int{\frac{dN}{dz}}dz=1$, which can be easily achieved
numerically.

\subsection{Estimating $dN/dz$}

To determine the correlation length, we need to estimate $dN/dz$.  Given a
photometric redshift and its random error, there is a conditional probability
of having a galaxy at a certain true redshift.  In addition, we need to
incorporate the apparent magnitude limit and photometric redshift selection
criteria in the estimate of the redshift distribution.  The real redshift $z$
and the photometric redshift $s$ of a galaxy are different. We assume that
$s=z+\nu$ where $\nu$ is the error and drawn from a normal distribution. Thus
\begin{equation}
P(s|z) = \frac{1}{\sqrt{2\pi\sigma^2}}
	 \exp\left\{{-\frac{(s-z)^2}{2\sigma^2}}\right\},
\label{eq:psz} 
\end{equation} 
where $\sigma$ determines the precision of the estimates. We need the
inverse: what is the true redshift, given the photometric estimate. This may
be obtained from Bayes' theorem,
\begin{equation}
P(z|s) = \frac{P(z)\, P(s|z)}{P(s)},
\end{equation}
where $P(z)$ is the true redshift distribution calculated from the LF that
also depends on the apparent magnitude cuts. 

Our volume limited sample is selected by a
window function of photometric redshifts,
\begin{equation}
W(s) = 
\begin{cases}
  1 & \text{if $s$ between 0.1 and 0.3,}\\
  0 & \text{otherwise.}\\
\end{cases}
\end{equation}
The conditional probability of having a galaxy in a sample selected by
this window function is calculated by the integral over the distribution
$P(s)$, 
\begin{align}
P(z|W)  & = \int\!\! ds\, P(s)\, W(s)\, P(z|s) \\
	& = P(z) \int\!\! ds\, W(s)\, P(s|z) \\
	& = P(z)\, W_{\sigma}(z), 
\end{align}
where $W_{\sigma}(z)$ is the photometric redshift selection function
convolved with the photometric redshift uncertainty.

\begin{figure}
\plotone{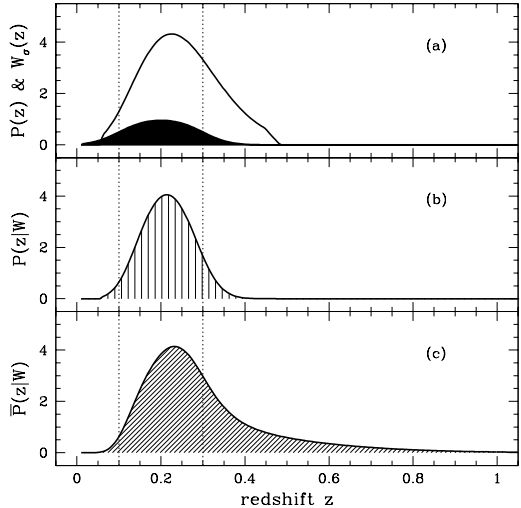}
\caption{The redshift distribution is computed by adding the contributions of
narrow $r^*$ intervals. The top panel (a) shows the number density of
galaxies with $r^*$ between 18 and 18.2 as function of redshift, as derived
from the luminosity function. The black shaded curve is the corresponding
window function with $\sigma=0.05$. The product of these two, shown in panel
(b), is the effective contribution of the magnitude bin.  Panel (c)
illustrates the final redshift histogram, which is the properly weighted sum
of the products like the one above.}
\label{fig:dndz}
\end{figure}

The precision of photometric redshifts is a strong function of the
apparent magnitude.  From the photometric redshift catalog, we compute the
mean redshift errors $\sigma_i$ and the galaxy counts $n_i$ (for proper
weighting) in $\Delta{}r^*=0.2$ wide magnitude bins. Using the LF by
\citet{blanton03}, we derive the redshift distributions $P_i(z)$ for the same
magnitude bins up to $r^*=21$, which is the limiting magnitude in the sample.
The final redshift distribution is the weighted average of these
probabilities,
\begin{equation}
\bar{P}(z|W) \propto \sum_i n_i\, P_i(z)\, W_{\sigma_i}(z) .
\end{equation}
The top panel of Figure~\ref{fig:dndz} illustrates the redshift
distribution derived from the LF for galaxies with
$18\!<\!\!r^*\!\!<\!\!18.2$ and a smoothed window $W_{\sigma}(z)$ with
$\sigma=0.05$. The middle panel shows the effective contribution of this
magnitude bin to the final $dN/dz$, which is plotted in the last
panel.

\subsection{Results for $r_0$}

Applying the above probabilistic redshift distribution and substituting the
power-law fits derived earlier, we find a correlation length for the full
volume limited sample of $5.77\,h^{-1}\text{Mpc}$. Estimating the uncertainty
of this value is not trivial. The statistical errors on the parametric fits
to $w(\theta)$ are known and
may be used to estimate the errors on the correlation length.
These are estimated by calculating the scatter of the
predicted $r_0$ measurements for 20,000 Monte-Carlo realizations of the
fitting parameters, $A_w$ and $\delta$, based on their covariance matrix. For
the full volume limited sample, we obtain
$\Delta_{\text{rms}}=0.05\,h^{-1}\text{Mpc}$.
However, the uncertainty on $r_0$ is also affected by the uncertainty in the
redshift distribution.  The primary source of change in the $dN/dz$ is the
uncertainty in the LF parameters. We compute the partial derivatives
$(\partial{}r_0/\partial{}\alpha)$,
$(\partial{}r_0/\partial{}M_*)$ and 
$(\partial{}r_0/\partial{}Q)$
numerically and propagate the quoted errors of \citet{blanton03}.  We find
that the evolutionary parameter $Q$ makes the largest difference, the $r_0$
errors from the uncertainty of $M_*$ and $\alpha$ are negligible. For the
fiducial $r_0$ value, we estimate an error of
$\Delta_{\text{LF}}=0.09\,h^{-1}\text{Mpc}$.
The dependence of $\sigma$ on the apparent magnitude was determined
empirically using the actual measurements, thus the results are not affected by
Malmquist bias.

\begin{figure*}
\plottwo{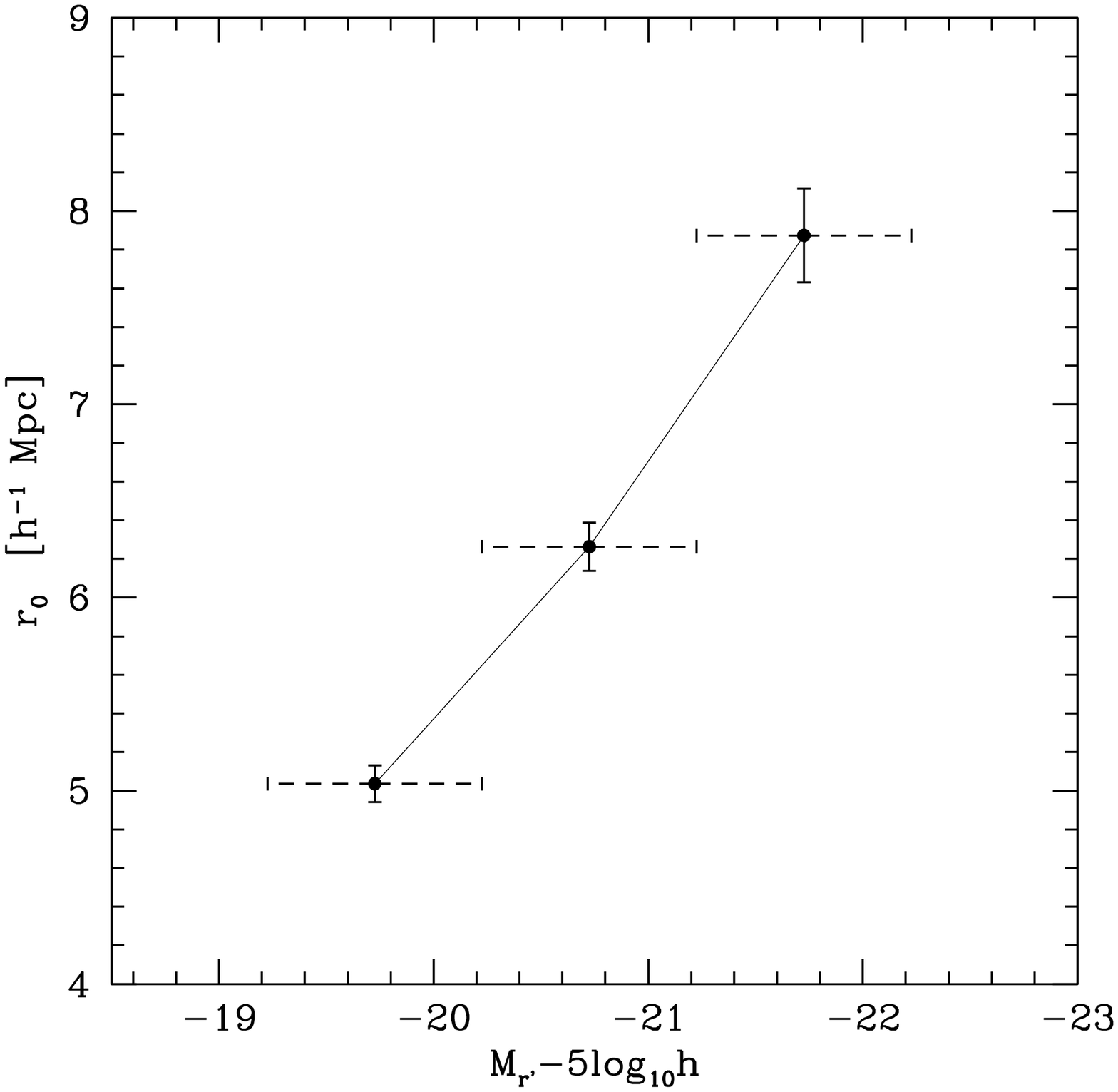}{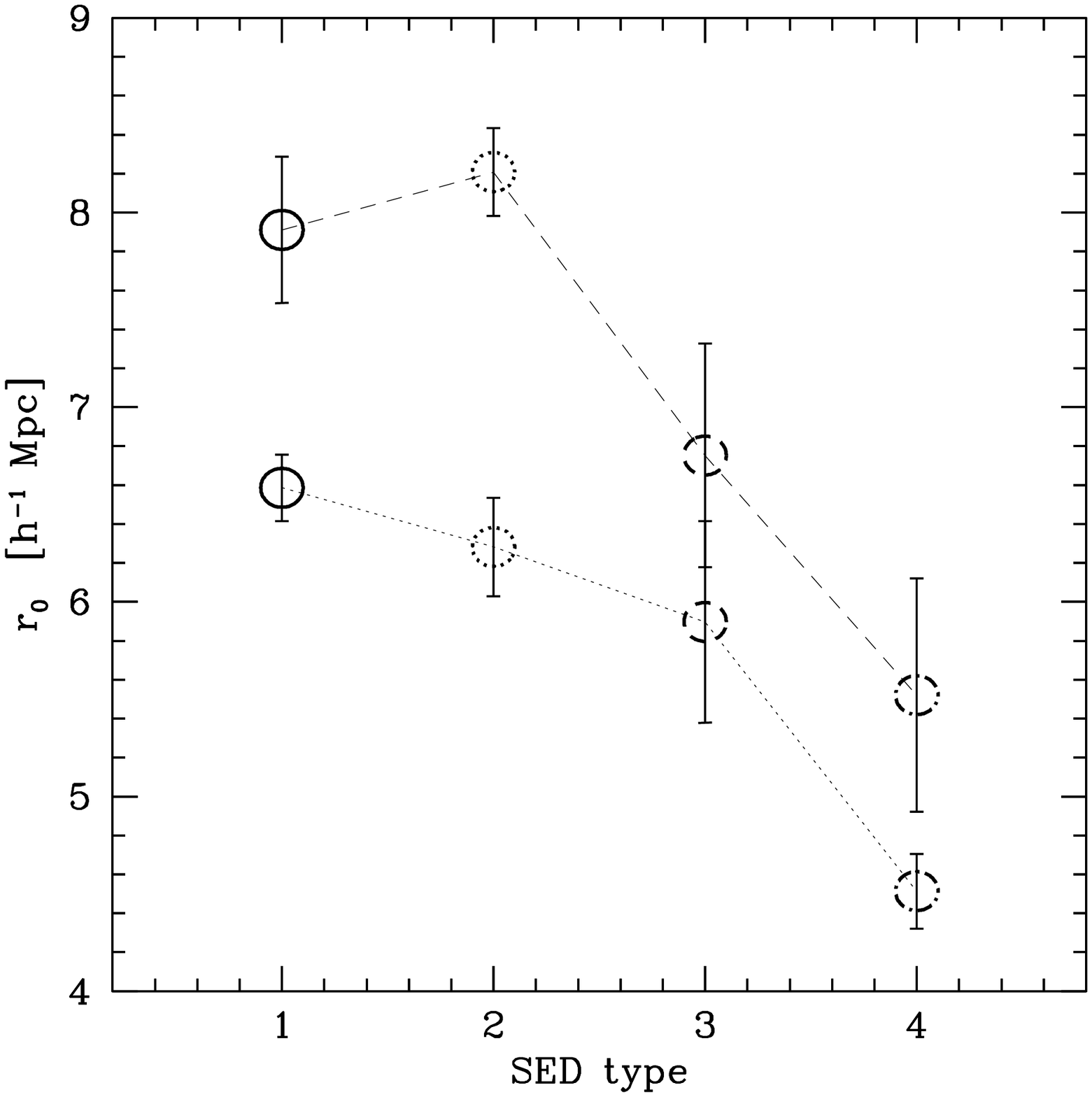}
\caption{The correlation length $r_0$ is plotted as a function of luminosity
({\it{}left}) and spectral type ({\it{}right panel}). The error bars include
both error terms added in quadrature.
\label{fig:rnut}}
\end{figure*}

In Figure~\ref{fig:rnut}, the correlation length is plotted as a function of
luminosity ({\em left panel}) and SED type ({\em right panel}). The relation
between $r_0$ and luminosity and spectral type are consistent with that
observed directly from the angular data and from measures of the clustering
length from spectroscopic surveys \citep{zehavi02}. Over an absolute
magnitude range of $-19.97$ to $-22$, $r_0$ increases with luminosity from a
value of $5.04\pm{}0.09$ to $7.87\pm{}0.24\,h^{-1}\text{Mpc}$, which is
consistent with the increase observed by \citet{zehavi03}.

The color dependence of $r_0$ (i.e.\ clustering as a function of spectral
type) shows the expected increase in clustering length for early type
galaxies.  The values of the correlation length for the spectral type
subsamples are given in Table~\ref{tbl:all}. In the lower luminosity bin red,
$T_1$, galaxies have a correlation length of
$6.59\pm{}0.17\,h^{-1}\text{Mpc}$ for the $M_{r^*} > -21$ sample and the
bluest galaxies, $T_4$, have a correlation length of
$4.51\pm{}0.19\,h^{-1}\text{Mpc}$. The trend in this relation is again
consistent with the observed dependence of the correlation length as a
function of spectral and morphological type \citep{giovanelli}.

\section{Discussion} \label{sec:concl}

The interesting aspect of the luminosity and color dependent
clustering is not that the observed clustering length scales with luminosity
and color as this has been demonstrated from many different
surveys. It arises from how the shape of the correlation function
depends on luminosity and color. It is remarkable that the luminosity
simply effects the amplitude of the correlation function and not the
slope, whereas the type selection affects both the slope and
amplitude. This is particularly intriguing when we note that we would
expect an intrinsic correlation between the luminosity and spectral
type of a galaxy.

Observationally, early type galaxies tend to reside in clusters of galaxies
whereas later type galaxies are more often found in the field. We would
expect, therefore, that early type galaxy samples would have more small
separation pairs than samples selected for late type galaxies and that their
resulting correlation functions would be steeper. This is consistent with the
data except that we would expect there to be a smooth transition from early
to late type galaxies and that the correlation function slope should smoothly
change from a steep value of $0.96$ for the early types to the more shallow
value of $0.68$ for the late types. What we observe, however, is that red
galaxies (types $T_1$ and $T_2$) have a common slope of $\sim 0.1$ and blue
galaxies (types $T_3$ and $T_4$) have a slope of $0.07$ (i.e.\ there does not
appear to be a smooth transition).

We can, however, explain this behavior if we consider a simple model for the
distribution of galaxy types. From Figure~\ref{fig:typehist} we see that the
type histogram for the galaxies is almost bimodal \citep{strateva01,hogg03}
with the 
distribution being well fit by two Gaussians. For simplicity we will denote
these subclasses as ``red'' and ``blue''. If the ``red'' and ``blue''
populations have distinct correlation functions (i.e.\ with different slopes),
then any observed correlation function should simply come from mixing these
populations. As we change the mix of ``red'' and ``blue'' galaxies then the
resulting slope of the correlation function will also change. This is exactly
what we observe with the correlation function as we move from the $T_2$ to
$T_3$ selections.

If we selected galaxies only from either the ``red'' or ``blue''
sub-populations we would expect no change in the correlation function slope
as all of the ``red'' or ``blue'' galaxies have a common correlation
function. Again, this what we observe from the data. The slopes of the
correlation functions for the $T_1$ and $T_2$ red samples are identical as
are the correlation functions for the blue $T_3$ and $T_4$ samples. In
reality, the color selection that we applied to the SDSS data (types $T_1$
through $T_4$) was not chosen to optimally separate two distinct populations
of galaxies but rather to provide a simple subdivision of galaxies based on
the CWW spectral energy distributions.  We might expect there to remain some
population mixing in our $T_1$, $T_2$, $T_3$ and $T_4$ color cuts. We observe
this effect where the amplitude of the correlation function for the $T_3$ and
$T_4$ classes are close but not identical; this would imply that the $T_3$
still contains a subset of the ``red'' galaxies.

The luminosity dependence can be explained if we note that the luminosity
functions of the ``red'' and ``blue'' classes are identical for magnitudes
brighter than $M_{r^*}=-20$ \citep{baldry}. They deviate only for the faint
end of the luminosity function (``blue'' galaxies having a steeper faint end
slope). Varying the luminosity cuts should not change the mix of the galaxy
populations (unless we sample galaxies with $M_{r^*}>-20$). We would,
therefore, expect the shape of the correlation function to be independent of
the luminosity cuts (as is found from the data). Given this hypothesis if we
selected a volume limited sample for galaxies with luminosities less that
$M_{r^*}=-20$ we would expect to find a dependence on the slope with
luminosity.

It is, therefore, remarkable that with such a simple model for the
distribution of galaxies (i.e.\ just two classes with differing
correlation functions) we can qualitatively describe the behavior of
the correlation functions with color and luminosity. What is
difficult to understand is why there would be a simple scaling of the
amplitude of the correlation function with intrinsic luminosity as the
spatial scales we sample are in the non-linear regime (i.e.\ a simple
linear bias model is not necessarily appropriate). Identifying the
physical mechanism that could give rise to the luminosity and type
dependent bias that we observe remains an open question.

\acknowledgments

We would like to thank Michael Strauss and Michael Blanton for their valuable
comments. AS was supported by NSF grants AST-9802 980, KDI/PHY-9980044, NASA
LTSA NAG-53503 and NASA NAG-58590. TB acknowledges support from NSF
KDI/PHY-9980044. IC, AS and TB acknowledge partial support from MTA-NSF
grant no.\ 124 and the Hungarian National Scientific Reasearch Foundation
(OTKA) grant T030836. AJC acknowledges partial support from an NSF career
award AST99-84924 an ITR NSF award 0121671. RS acknowledges partial support
form ITR NSF award 0121671. IS was supported by NASA through AISR grants
NAG5-10750, NAG5-11996, and ATP grant NASA NAG5-12101 as well as by NSF
grants AST02-06243.

Funding for the creation and distribution of the SDSS Archive has been
provided by the Alfred P.\ Sloan Foundation, the Participating
Institutions, the National Aeronautics and Space Administration, the
National Science Foundation, the U.S. Department of Energy, the
Japanese Monbukagakusho, and the Max Planck Society. The SDSS Web site
is \url{http://www.sdss.org/}.

The SDSS is managed by the Astrophysical Research Consortium (ARC) for the
Participating Institutions. The Participating Institutions are The University
of Chicago, Fermilab, the Institute for Advanced Study, the Japan
Participation Group, The Johns Hopkins University, Los Alamos National
Laboratory, the Max-Planck-Institute for Astronomy (MPIA), the
Max-Planck-Institute for Astrophysics (MPA), New Mexico State University,
Princeton University, the United States Naval Observatory, University of
Pittsburgh and the University of Washington.

\begin{deluxetable}{lccrccc}
\tablecolumns{6}
\tablewidth{0pc}  
\tablecaption{Power-law fits to correlation functions and correlation lengths}
\tablehead{
\colhead{Sample} & 
\colhead{Luminosity} &
\colhead{SED type} &
\colhead{$N$\tablenotemark{\dag}} & 
\colhead{$A_w$} & 
\colhead{$\delta$} &
\colhead{$r_0$\tablenotemark{\ddag}}}
\startdata
Fiducial & All & All & 2,016 & $0.078\pm{}0.001$ & $0.84\pm{}0.02$ & $5.77\pm{}0.05\pm{}0.09$ \\
\cline{1-7}
I & $M_{r^*}\!>\!-21$ & All & 1,098 & $0.061\pm{}0.001$ & $0.84\pm{}0.03$ &$5.04\pm{}0.05\pm{}0.08$ \\
 & $-21\!>\!M_{r^*}\!>\!-22$ & All & 650 & $0.092\pm{}0.002$ & $0.85\pm{}0.04$ & $6.26\pm{}0.08\pm{}0.10$ \\
 & $-22\!>\!M_{r^*}\!>\!-23$ & All & 268 & $0.138\pm{}0.004$ & $0.84\pm{}0.05$ & $7.87\pm{}0.21\pm{}0.12$ \\
\cline{1-7}
II & $M_{r^*}\!>\!-21$ & $t<0.02$ & 343 & $0.112\pm{}0.004$ & $0.96\pm{}0.05$ & $6.59\pm{}0.14\pm{}0.10$ \\
 & $M_{r^*}\!>\!-21$ & $0.02<t<0.3$ & 254 & $0.109\pm{}0.005$ & $1.02\pm{}0.06$ & $6.28\pm{}0.24\pm{}0.09$ \\
 & $M_{r^*}\!>\!-21$ & $0.3<t<0.65$ & 185 & $0.072\pm{}0.008$ & $0.66\pm{}0.12$ & $5.90\pm{}0.51\pm{}0.10$ \\
 & $M_{r^*}\!>\!-21$ & $t>0.65$ & 316 & $0.047\pm{}0.003$ & $0.68\pm{}0.09$ & $4.51\pm{}0.18\pm{}0.07$ \\
\cline{1-7}
III & $-21\!>\!M_{r^*}\!>\!-23$ & $t<0.02$ & 280 & $0.161\pm{}0.007$ & $0.96\pm{}0.05$ & $7.91\pm{}0.36\pm{}0.12$ \\
 & $-21\!>\!M_{r^*}\!>\!-23$ & $0.02<t<0.3$ & 326 & $0.155\pm{}0.005$ & $0.87\pm{}0.04$ & $8.21\pm{}0.19\pm{}0.12$ \\
 & $-21\!>\!M_{r^*}\!>\!-23$ & $0.3<t<0.65$ & 185 & $0.093\pm{}0.010$ & $0.70\pm{}0.15$ & $6.75\pm{}0.56\pm{}0.11$ \\
 & $-21\!>\!M_{r^*}\!>\!-23$ & $t>0.65$ & 127 & $0.067\pm{}0.010$ & $0.72\pm{}0.17$ & $5.52\pm{}0.59\pm{}0.09$ \\
\enddata
\tablenotetext{\dag}{Number of galaxies in subsample ($\times 10^3$)}
\tablenotetext{\ddag}{Correlation length in $h^{-1}$Mpc. The two estimates
are the statistical error from the power-law fits and the 
error from the uncertainty of the luminosity function parameters.
\label{tbl:all}}
\end{deluxetable}


\begin{thebibliography}{99}
\bibitem[DR1; Abazajian {\etal} (2003)]{dr1} Abazajian, K., \etal, 2003, in 
  preparation 

\bibitem[Arnouts \etal (1999)]{1999MNRAS.310..540A} Arnouts, S., Cristiani,
  S., Moscardini, L., Matarrese, S., Lucchin, F., Fontana, A., \& Giallongo,
  E.\ 1999, \mnras, 310, 540 

\bibitem[Baldry \etal (2003)]{baldry} Baldry, I.K., \etal, 2003, in
  preparation

\bibitem[Baugh \etal (1999)]{1999MNRAS.305..L21} Baugh, C.~M., Benson, A.~J.,
  Cole, S., Frenk, C.~S., \& Lacey, C.~G.\ 1999, \mnras, 305, L21
  
\bibitem[Blanton \etal (2003)]{blanton03} Blanton, M.R., \etal, 2003, \apj, 
  in press

\bibitem[Brent(1973)]{brent} Brent, R.P., 1973, Algorithms for
	Minimization without Derivatives (Englewood Cliffs, NJ:
	Prentice-Hall), Chapter 5.

\bibitem[Brunner, Szalay \& Connolly(2000)]{brunner00} Brunner, R.J., Szalay,
	A.S., Connolly, A.J., 2000, \apj, 541, 527

\bibitem[Budav\'ari \etal (1999)]{budavari99} Budav\'ari, T., Szalay, A.S.,
        Connolly, A.J., Csabai, I., \& Dickinson, M.E., 1999, in 
        {\em Photometric Redshifts and High Redshift Galaxies,}  eds.\ R.J.\
        Weymann, L.J.\ Storrie--Lombardi, M.\ Sawicki, \& R.\ Brunner, (San
        Francisco:  ASP), 19

\bibitem[Budav\'ari \etal (2000)]{budavari00} Budav\'ari, T., Szalay, A.S.,
        Connolly, A.J., Csabai, I., \& Dickinson, M.E., 2000, \aj, 120, 1588

\bibitem[Coleman, Wu \& Weedman(1980)]{cww} Coleman, G.D., Wu., C.-C., \&
        Weedman, D.W., 1980, \apjs, 43, 393 

\bibitem[Connolly \etal (1995a)]{connolly95a} Connolly, A.J., Csabai, I.,
        Szalay, A.S., Koo, D.C., Kron, R.G., \& Munn, J.A., 1995a, \aj, 110,
        2655

\bibitem[Connolly \etal (1998)]{connolly98} Connolly, A.J., Szalay, A.S.,
	Brunner, R.J., 1998, \apj, 499, L125

\bibitem[Connolly \etal (1999)]{connolly99} Connolly, A.J., Budav\'ari, T.,
        Szalay, A.S., Csabai, I., \& Brunner, R.J., 1999, in 
        {\it Photometric Redshifts and High Redshift Galaxies,}  eds.\ R.J.\
        Weymann, L.J.\ Storrie--Lombardi, M.\ Sawicki, \& R.\ Brunner, (San
        Francisco:  ASP), 13

\bibitem[Csabai \etal (2000)]{csabai00} Csabai, I., Connolly, A.J., Szalay,
        A.S., \& Bu\-da\-v\'{a}\-ri, T., 2000, \aj, 119, 69

\bibitem[Csabai \etal (2002)]{csabai02} Csabai, I., \etal, 2002, 
	accepted to \aj

\bibitem[Firth \etal (2002)]{2002MNRAS.332..617F} Firth, A.~E.~et al.\ 
2002, \mnras, 332, 617 

\bibitem[Fukugita \etal (1996)]{fukugita} Fukugita, M., Ichikawa, T., Gunn,
	J.E., Doi, M., Shimasaku, K. \& Schneider, D.P. 1996, \aj, 111, 1748 

\bibitem[Giovanelli \etal (1986)]{giovanelli} Giovanelli, R., Haynes, M.P.,
	Chincarini, G.L., 1986, \apj, 300, 77

\bibitem[Gunn \etal (1998)]{gunn} Gunn, J.E., Carr, M.A., Rockosi, C.M.,
	Sekiguchi, M., et al. 1998, AJ, 116, 3040

\bibitem[Hamilton(1993)]{hamilton} Hamilton, A.J.S., 1993, \apj, 417, 19

\bibitem[Hogg \etal (2001)]{hogg01} Hogg, D.W., Finkbeiner, D.P., Schlegel,
	D.J., \and Gunn, J.E., 2001, \aj, 122, 2129 

\bibitem[Hogg \etal (2003)]{hogg03} Hogg, D.W., \etal, 2003, \apjl, 
  	in press 

\bibitem[Kerscher \etal (2000)]{kerscher00} Kerscher, M., Szapudi, I.,
	Szalay, A.S., 2000, \apj, 535, 13

\bibitem[Landy \& Szalay(1993)]{landyszalay} Landy, S.D., \& Szalay, A.S.,
	1993, \apj, 412, 64

\bibitem[Lupton \etal (2003)]{lupton} Lupton, R.H., \etal, 2003, in preparation

\bibitem[Magliocchetti \& Maddox(1999)]{maglio99} Magliocchetti, M., \& Maddox,
	S.J., 1999, \mnras, 306, 988

\bibitem[Norberg \etal (2001)]{norberg01} Norberg, P., \etal, 2001, \mnras,
	328, 64

\bibitem[Peebles(1980)]{peebles} Peebles, P.J.E., 1980, in {\em Large-Scale
	Structure of the Universe}, 174

\bibitem[Pier \etal (2002)]{pier02} Pier, J.R., \etal, 2002, \aj, in press,
  	astro-ph/0211375 

\bibitem[Press \etal (1992)]{nr} Press, W.H., Teukolsky, S.A., Vetterling,
	W.T., \& Flannery, B.P., 1992, in {\em Numerical Recipes in C}
	(Cambridge University Press)

\bibitem[Roukema \etal (1999)]{1999MNRAS.305..151R} Roukema, B.~F.,
  Valls-Gabaud, D., Mobasher, B., \& Bajtlik, S.\ 1999, \mnras, 305, 151 

\bibitem[Scranton \etal (2003)]{scran03} Scranton, R., \etal, 2003, in
	preparation 

\bibitem[Smith \etal (2002)]{smith02} Smith, J.A., \etal, 2002, \aj, 123,
  	2121

\bibitem[Stoughton \etal (2002)]{stoughton} Stoughton, C., \etal, 2002, \aj,
	123, 485

\bibitem[Strateva \etal (2001)]{strateva01} Strateva, I., \etal, 2001, \aj, 
	122, 186

\bibitem[Szapudi, Prunet \& Colombi(2001)]{spice} Szapudi, I., Prunet, S., \&
	Colombi, S., 2001, \apj, 561, 11

\bibitem[Szapudi \etal (2003)]{espice} Szapudi, I., \etal, 2003, in
	preparation (eSpICE)

\bibitem[Teplitz \etal (2001)]{2001ApJ...548..127T} Teplitz, H.~I., Hill, 
R.~S., Malumuth, E.~M., Collins, N.~R., Gardner, J.~P., Palunas, P., \& 
Woodgate, B.~E.\ 2001, \apj, 548, 127 

\bibitem[Thakar \etal (2001)]{sx} Thakar, A.R., Kunszt, P.Z., Szalay, A.S.,
	2001, in {\em Mining the Sky}, eds.\ A.J.\ Banday et al., (Garching:
	ESO Astrophysics Symposia, Proceedings 2000.\ XV), 624

\bibitem[Totsuji \& Kihara(1969)]{totsujikihara} Totsuji, H., \& Kihara, T., 
  	1969, \pasj, 21, 221

\bibitem[SDSS; York \etal (2000)]{york} York, D.G., \etal, 2000, \aj, 120,
        1579

\bibitem[Zehavi \etal (2002)]{zehavi02} Zehavi, I., \etal, 2002, \apj, 
	571, 172

\bibitem[Zehavi \etal (2003)]{zehavi03} Zehavi, I., \etal, 2003, 
	in preparation

\end{thebibliography}
\end{document}